\documentclass[aps,prl,showpacs,twocolumn,superscriptaddress,english]{revtex4-1}

\usepackage{graphicx}
\usepackage{amsmath}%
\usepackage{color}

\usepackage{lipsum, babel}

\usepackage{hyperref}

\hypersetup{
    colorlinks=true,
    linkcolor=blue,
    filecolor=magenta,      
    urlcolor=cyan,
}
 \usepackage{url}
\begin{document}

\title{Gapped electron fractionalization in robustly one dimensional Li$_{0.9}$Mo$_6$O$_{17}$}

\author{Natalia Lera}
\affiliation{Departamento de F\'isica de la Materia Condensada, Condensed Matter Physics Center (IFIMAC) and Instituto Nicol\'as Cabrera, Universidad Aut\'onoma de Madrid, Madrid 28049, Spain}

\author{J.V. Alvarez} 
\affiliation{Departamento de F\'isica de la Materia Condensada, Condensed Matter Physics Center (IFIMAC) and Instituto Nicol\'as Cabrera, Universidad Aut\'onoma de Madrid, Madrid 28049, Spain}
 
 \author{J. W. Allen}
 \affiliation{Randall Laboratory of Physics, University of Michigan, Ann Arbor, Michigan 48109, USA}

\begin{abstract}
The angle resolved photoemission spectroscopy lineshapes of quasi one-dimensional (1d) Li$_{0.9}$Mo$_6$O$_{17}$ display both agreement with and departures from the one-band Tomonaga-Luttinger model. We show that the departures can be understood by explicitly accounting for the \emph{four} modes arising from the \emph{two} quasi-1d bands known to cross the Fermi energy. The key assumption is that the antisymmetric charge mode is gapped with a magnitude near the temperature (T) of a mysterious 25K powerlaw resistivity upturn. The gap is consistent with the lack of a charge or spin density wave accompanying the upturn, is able to control the upturn T, and prevents crossover to a Fermi liquid (FL) down to the superconducting transition at 1.9K.

\end{abstract}

\maketitle

In a quantum critical (QC)  metal, the thermal coherence length $\lambda_T$ is the only scale defined in any correlation function. Experimentally, this scaling manifests itself in the universal dependencies of spectral densities and susceptibilities on frequency $\omega$ and momentum $k$ as $\omega/T$ and $k/T$. As T goes to zero and $\lambda_T$  diverges, these dependencies simplify to power laws, whose critical exponents are not independent but are related by strict scaling relations. An archtype example of QC behavior is the Luttinger liquid (LL) properties of the one-band one-dimensional (1d) Tomonaga Luttinger (TL) model \cite{Tomonaga1950,Luttinger1963}.  The detailed elucidation of QC behavior in complex quantum materials requires the comparison of excellent spectroscopy with an accurate spectral theory. These conditions are met in quasi-1d Li$_{0.9}$Mo$_6$O$_{17}$, the so-called lithium purple bronze (LiPB).  Except that LiPB has two bands crossing the Fermi energy $E_F$, the material fulfills the basic provisos for considering the one-band TL model fixed point as a starting point for analysis and modelling: i) a quasi-one-dimensional Fermi surface \cite{gweon2003}, ii) linearity of the bands crossing $E_F$ over more than 0.1eV \cite{JV2006}, iii) large value of the density of states exponent $\alpha \sim 0.6$ \cite{JV2006,cazalilla2005,Matzdorf2013} which partially protects the system against single-particle perturbations and simultaneously suggests, iv) domination of  forward-scattering interactions. 

	A series of angle resolved photoemission spectroscopy (ARPES) experiments \cite{denlinger,gweon2001,gweon2002,gweon2003,gweon2004,SolidStateAllen,JV2006,JV2009,Allen13}, scanning tunnel spectroscopy (STS) \cite{cazalilla2005,Matzdorf2013} and transport \cite{hussey2011} show simultaneously quantitative agreements with, and clear-cut deviations from, a Luttinger Liquid (LL) metal \cite{Tomonaga1950,Luttinger1963, Haldane1981,GiamarchiBook,Chudzinski}.  The critical exponents are well defined and have been extracted experimentally in a consistent manner using a variety of experimental techniques and analysis procedures. In the absence of a two-band theory, the analysis has proceeded by comparison with single-band TL spectral densities \cite{Orgad01}.  In our discussion below we will define $\alpha_{\rm PL}$ as the exponent of the density of states (DOS) extracted from k-integrated  photoemission spectroscopy (PES) using a fit to the single-band Luttinger Liquid (LL) \cite{JV2006,JV2009}, and  $\alpha_{\rm STS}$  the equivalent exponent measured with STS \cite{cazalilla2005,Matzdorf2013}.  We define $\eta_{\rm PL}$ as the exponent of the ARPES energy distribution curves (EDC) temperature-prefactor, obtained by scaling normalized T-dependent EDC data for $k=k_F$ to match at $\omega=E_F$. The experimental findings of Ref. \cite{Allen13} are:  $\alpha_{\rm PL} \approx 0.9$ at high temperatures ($T=300K$), $\alpha_{\rm PL} \approx 0.6$ at lower temperatures ($T \leq 150K$),  consistent with STS measurements \cite{STMLiPB} $\alpha_{\rm STS}\approx 0.6$, and $\eta_{\rm PL} \approx 0.6$.  These values suggest  $\alpha_{\rm PL} \simeq \eta_{\rm PL}$, reported for the first time in Ref.\cite{JV2009}.  Although all this extensive and consistent experimental  analysis used, quite succesfully, analytical  TL spectral densities to fit the spectra, there is a basic inconsistency: the TL theory implies necessarily the scaling law $\alpha_{\rm TL}=\eta_{\rm TL}+1$.  In further disagreement, only the spinon edge of the spectral function shows this QC scaling, but not the holon peak.  It is natural to consider whether these disagreements could arise from the two-band nature of LiPB.

In this Letter we construct a rigorous spectral theory that takes into account that a two-band TL liquid has four boson  modes (two charge and two spin) rather than two, with the key assumption that the new asymmetric charge mode is gapped, and we compare with all the available spectroscopy for the material. We can show why the LL scaling law $\alpha= \eta+1$ breaks down, we can interpret the scaling of the spinon edge, and we can provide a microscopic origin for a phenomenological description \cite{Allen13} of the imperfect scaling of the EDC's. Our spectral densities are in good agreement with the observed EDC's.
 
Although QC scaling is our main focus here, we also point out the likely role of non-critical fluctuations in the gapped charge mode, for the unusually robust 1-d metallic behavior \citep{Allen13}, the mysterious resistivity upturn at 20-30K, and the superconductivity proposed to be triplet \cite{mercure2012,lebed2013_1,raghu,nl1}.

 Ref. \cite{Allen13} presents a successful phenomenological formula to fit the experimental ARPES results, based on the single-band TL spectral function modifed by the following convolution: 

\begin{equation}
A_{test}(k,\omega;T)=T^{\eta_{TL}+1}\int_{-\infty }^{\infty } R\left (\frac{p}{p_0} \right ) A_{\rm TL}(\tilde{k}-\tilde{p},\tilde{\omega})d\tilde{p}
\label{phenomenological}
\end{equation}

\begin{figure}
  \includegraphics[width=40mm,height=40mm]{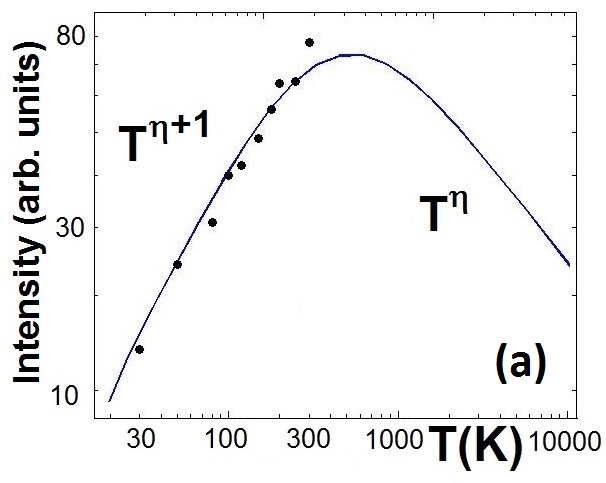}
  \includegraphics[width=40mm,height=40mm]{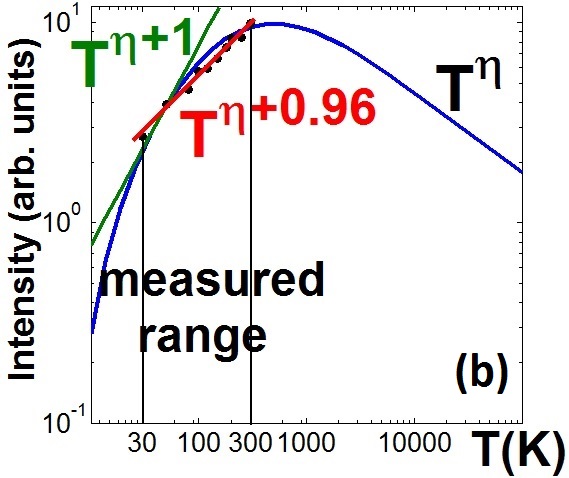}
\caption {(Color online) Log-log plot of the spectral function  at $E_F$ and $k_F$ against temperature. The black dots are the experimental results \cite{Allen13}
(a) single LL convolved with a Gaussian. 
(b) gapped theory.  Both approaches show a power law $T^{\eta}$ behavior at the measured range of temperatures. Here, $\eta$ refers to the experimental exponent: (a) $\eta_{\rm PL}$ (b) $\eta_{\rm PG}$
}
\label{etaExp}
\end{figure}

R is a broad normalized Gaussian $R\left (\frac{p}{p_0} \right )=\frac{1}{\sqrt{\pi}p_0}\exp\left (-p^2/p_0^2\right )$ with $p_0=0.065 \AA^{-1}$ and A is the single-band Tomonaga-Luttinger (TL) spectral function with arguments $\tilde{k}=k\lambda_\sigma$, $\lambda_\sigma=\frac{v_\sigma}{\pi T}$ and $\tilde{\omega}=\frac{\omega}{\pi T}$, and parameters $\frac{v_c}{v_\sigma}=2$, $v_\sigma=1.9 eV \AA$ and $\eta_{TL}+1=0.6$. Notice that $\eta_{TL}$ apart from being the T-exponent is also related to an interaction parameter $\beta_{\rho}$ (defined below) in a single band LL with spin rotational invariance, $\eta_{TL}=2\beta_{\rho}-1$, and this dependence is included in the calculation of $A_{\rm TL}$. Also notice that, apart from the temperature prefactor, the momentum of the $A_{test}$ function does not scale with temperature; nevertheless we still can define an $\eta_{\rm PL}$ exponent carrying the T-dependence of the scaled Green's function, observable in EDC lineshapes. The $\eta_{\rm PL}$ exponent will depend on temperature as shown in Fig. \ref{etaExp}(a). PES experiments measuring up to 300K extract a value of $\eta_{PL}\approx 0.6$ inconsistent with LL theory $\eta_{TL}+1=0.6$.

After the convolution with the Gaussian in the high temperature limit we have $\eta_{\rm PL}=\eta_{\rm TL}$, while in the range of temperatures of interest, the observed exponent is $\eta_{\rm PL}=\eta_{\rm TL}+1$ consistent with experiment, (see Fig. \ref{etaExp}(a)), as pointed out in Ref. \citep{Allen13}.

The values of $p_0$ and $\alpha_{\rm TL}$ determine the border between the two regimes. Keeping the idea of convolution in mind, we look for plausible models explaining the good agreement of Eq.(\ref{phenomenological}) with the experiment.\\

Both photoemission experiments and band theory calculations
\cite{Whangbo88,POPOVIC05,Nuss} show two nearly degenerate bands crossing $E_F$, and we now show that this offers a natural starting point for the phenomenological approach \citep{Allen13}.
Following the standard bosonization process we can decouple the
fixed-point Hamiltonian with two bands, in spin and charge modes and in the sum and difference of density fluctuations between the two bands:

\begin{equation}
H= \sum_{\substack{\mu=\rho,\sigma \\ \nu=\pm}}
\frac{v_{\mu \nu}}{2} \int dx \left[K_{\mu \nu}
(\Pi(x)_{\mu \nu})^2+ \frac{1}{K_{\mu \nu}} (\partial_x
\phi(x)_{\mu \nu})^2 \right] \nonumber
\end{equation}
where  $\Pi(x)_{\mu \nu}$ and $\phi(x)_{\mu \nu}$ are
the conjugate bosonic fields, $v_{\mu \nu}$ the sound
velocities and $K_{\mu \nu}$ the Luttinger parameters(
$\nu=+$/$-$ superposition/difference, $\mu=\rho$/$\sigma$
charge/spin density). 
The Green's Function relevant for ARPES spectra close to the Fermi momentum $k_F$, is $G_{<}(x,t;T)=\langle \Psi^{\dagger}(x,t) \Psi(0,0) \rangle $. It factorizes into a product of form factors coming from each independent mode.

\small
\begin{equation}
G_{<}(x,t;T)=\frac{e^{-iK_Fx}}{2\pi a}\prod_j f_j\left(\frac{x}{\lambda_j},\frac{v_jt}{\lambda_j};\beta_j\right)  \label{SMrsG}\\
\end{equation}
\normalsize
where j runs in the $N=4$ modes ($\rho+$, $\rho-$, $\sigma+$, $\sigma-$), $\lambda_j=\frac{ v_j}{\pi T}$ is the thermal coherence length, $\beta_j=\frac{1}{2N}(K_{\mu \nu}+K^{-1}_{\mu \nu}-2)$, $f_j$ is the form-factor for a gapless bosonic mode in 1d, and $a$ is a short-distance cut-off.

Assuming spin rotational invariance we have $K_{\sigma+}=K_{\sigma-}=1$ and thus $\beta_{\sigma+}=\beta_{\sigma-}=0$ .
The measured susceptibility is temperature independent down to temperatures close to the SC transition \cite{Chi_sGreenblatt,Chi_sMatsuda,Chi_sChoi}, and no magnetic instabilities have been observed in this range of temperatures. There are no other indications in the phenomenology of remarkable spin-dependence in the interactions.

 Therefore and for the sake of simplicity, we can assume that  $g^{(i)}_{j \parallel}=g^{(i)}_{j \perp}$  (i.e. $v_{\sigma+}=v_{\sigma-}=v_{\sigma}$). The resulting spin form factor is $f_{\sigma+}f_{\sigma-}=f_{\sigma}=\left[\left(\frac{\lambda_{\sigma}}{a}\right)h\left(\frac{x-v_{\sigma}t+ia}{\lambda_{\sigma}}\right)\right]^{-1/2}$ which is identical to the spin form factor of the conventional single-chain TL model.
We can treat the $\rho+$ mode as the charge mode in a single-band TL model, convolve charge and spin form factors and get an expression identical to the TL Green's function. We conclude that, under very general conditions the lesser Green's function of the two-band system can be written as
\begin{equation}
G_{<}(\tilde{k},\tilde{\omega})=\int \int d\tilde{q}d\tilde{\Omega} f_{\rho -}(\tilde{q},\tilde{\Omega})G_{<TL}\left ( \tilde{k}-\frac{\lambda_\sigma}{\lambda_{\rho -}}\tilde{q},\tilde{\omega}-\tilde{\Omega} \right )
\label{likeExperiment}
\end{equation}
i.e., a convolution of the $G_{<TL}$ with $f_{\rho -}$, the form factor of the extra $\rho-$ mode \cite{SM}. Notice that $G_{<}(\tilde{k},\tilde{\omega})$ scales in momentum and energy, the critical exponents are T-independent.\\ As we demonstrate in \cite{SM},
this model satisfies the TL scaling relation, $\eta+1=\alpha$ and it shows perfect scaling (as we show in \cite{SM} Fig. 1). Thus, it cannot account for the experimental observations.\\

To agree with the experimental results, we propose a different form factor for the $\rho -$ mode, that it acquires a gap. Since $\rho -$ represents the difference in charge between the two bands, this mode is expected to be small and chargeless as a first approximation, consistent with the lack \cite{Allen13} of charge density wave (CDW) signatures in LiPB. The gap is taken near the upturn in resistivity, i.e. $\Delta=30$K. The possibility of observing this small gap in experiment is discussed below.

A Luther-Emery form factor is proposed for $f_{\rho -}$:
\begin{equation}
f_{\rho -}(q,\Omega)= \left ( \frac{a}{\xi_{\rho -}}\right)^{\beta_{\rho -}+\frac{1}{4}} Z(q\xi_{\rho -})\delta(\Omega+E_{\rho -}(q))
\label{frhominusGap}
\end{equation}

This is actually the predominant spectral weight in a gapped 1D system belonging to the general class of the quantum sine-Gordon model \cite{Orgad01}. In such a case, one has $E_{\rho-}(q)=\sqrt{\Delta^{2}_{\rho-}+{(v_{\rho-}q)}^2}$,  $\xi_{\rho-}=\frac{v_{\rho-}}{\Delta_{\rho-}}$  and the correlation length $\xi_{\rho-}$ replaces the thermal coherence length 
$\frac{a}{\lambda_{\rho-}}\rightarrow \frac{a}{\xi_{\rho-}}$. The single free parameter of this theory is $\xi_{\rho-}$ or $v_{\rho-}$.
Multisoliton corrections to this single-mode form have a larger frequency threshold and very small weight.\\ 

Substituting Eq. (\ref{frhominusGap}) in Eq. (\ref{likeExperiment}) \cite{SM}, 
we get:
\begin{equation}
G_{<Gap}=cT^{\eta_{TG}}\int d\tilde{q}Z \left(\frac{\tilde{q}}{\xi_{\rho -}}\right)\frac{G_{<TL}(\tilde{k}-\frac{\lambda_\sigma}{\xi_{\rho -}}\tilde{q},\tilde{\omega}+\frac{E_{\rho -}}{\pi T})}{c_{TL}(T)} 
\label{EtaGapEq}
\end{equation}
	where $c$ is a constant, the ratio $G_{<TL}/c_{TL}(T)$ is function of the scaled variables ($\tilde{\omega}$ and $\tilde{k}$). Since $\rho -$ is non-critical and $\sigma$ is rotational invariant, $\eta_{TG}=2\beta_{\rho +}-\frac{5}{4}$ \cite{SM}. Based on experimental findings where $\alpha_{\rm PL}=0.6$, we set $\alpha_{\rm TG}=\alpha_{\rm PL}=0.6$ throughout the entire calculation so that $\beta_{\rho +}=0.425$ unequivocally. We will use this value of $\beta_{\rho +}$ throughout the article, meaning definite values of $\alpha_{\rm TG}$ and $\eta_{\rm TG}$ for all temperatures (the relation $\alpha_{\rm TG}=\eta_{\rm TG}+1$ still holds). The reader should not confuse theoretical values of $\alpha_{\rm TG}$ and $\eta_{\rm TG}$, defined in terms of interaction parameters, with the exponents $\alpha_{\rm PG}$ and $\eta_{\rm PG}$ extracted from a PES experiment using the gapped $\rho_{-}$ theory.
Within our theory we cannot give the same physical definition as for an LL because, at zero temperature, the energy argument of $G_{<TL}$ goes to $+ \infty$ and the intensity decay is larger than any power law. However, the definition in terms of interactions ($\beta_{\rho +}$) is similar.

 The correlation length $\xi_{\rho -}$, is T-independent and therefore, so is $\tilde{q}=q\xi_{\rho -}$. Since the argument of $Z$ does not scale, the integral is T-dependent and so is $\eta_{\rm PG}$, the observed exponent.\\

First we define a normalized $Z$ function, from \cite{Orgad01}:
\begin{equation}
Z(q)=\frac{v_{\rho -}}{\pi \Delta_{\rho -}} \left ( 1- \left ( \frac{v_{\rho -}q}{E_{\rho -}(q)}    \right )^2 \right )=\frac{\xi_{\rho-}/\pi}{1+\left ( \xi_{\rho-}q \right )^2}
\end{equation}

\begin{figure}
  
\includegraphics[width=80mm,height=40mm]{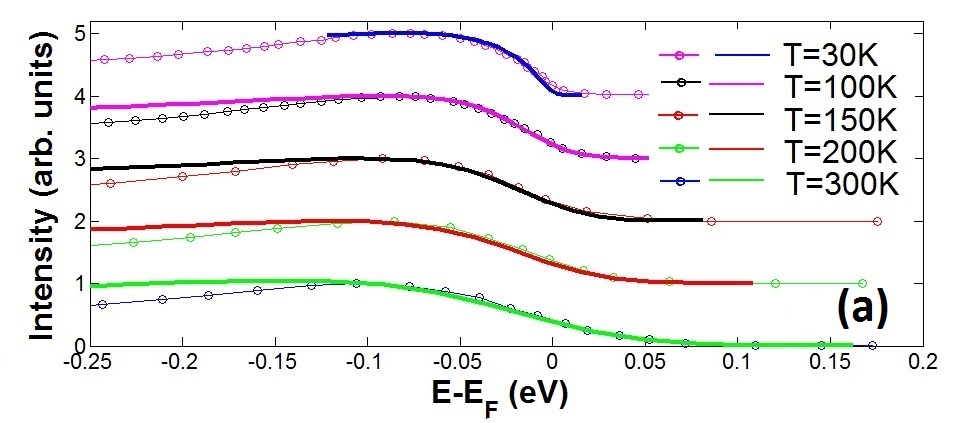}\\
\includegraphics[width=40mm,height=40mm]{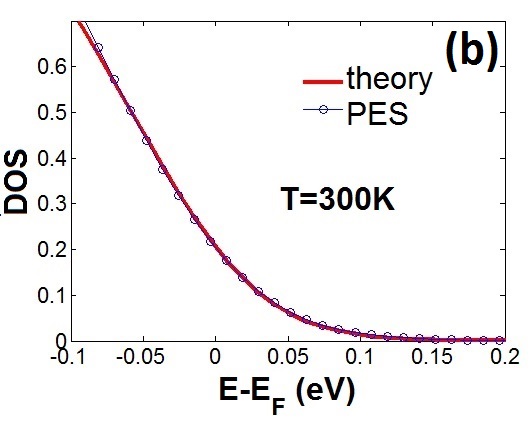}
\includegraphics[width=40mm,height=40mm]{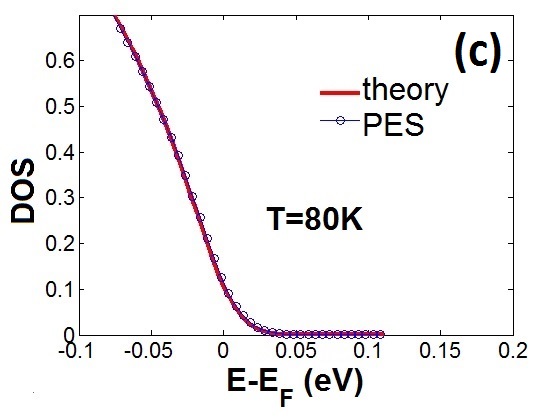}
    
\caption {(Color online) Comparison of gap mode theory (solid lines) to (a) previous fits of T-dependent EDC data using single-band TL phenomenology \cite{Allen13} and (b,c) previous single-band TL fits \citep{JV2006} of k-integrated data.}

\label{etaGap}
\end{figure}

The results for $\eta_{\rm PG}$ are shown in Fig. \ref{etaExp}(b). 
In the high temperature limit, $\frac{\lambda_\sigma}{\xi_{\rho -}}$ and $\frac{E_{\rho -}}{\pi T}$ go to zero, the integral in $\tilde{q}$ of Eq. (\ref{EtaGapEq}) gives the normalization of $Z$, and since $A_{<TL}$ scales, $\eta_{\rm PG}=\eta$. In the low temperature limit a change of variables $\tilde{q'}=\tilde{q}\lambda_\sigma$ and expanding $Z \left( \frac{\tilde{q'}}{\xi_{\rho -}}\right)$ in a Taylor series gives:
\begin{equation}
G_{<Gap}=cT^{\eta+1}\int d\tilde{q'}\frac{G_{<TL}(\tilde{k}-\frac{\tilde{q'}}{\xi_{\rho -}},\tilde{\omega}+\frac{E_{\rho -}}{\pi T})}{c_{TL}(T)} 
\end{equation}
where a constant has been included in a redefined $c$. 
In the measured range, the theory reproduces the observed exponent $\eta_{\rm PG}=\eta_{\rm TG}+1$, as shown in Fig. \ref{etaExp}(b). Fig. \ref{etaGap} shows that the theory also reproduces the 
previous good fits of k-integrated data and of experimental EDCs at $k=k_F$, based on the single-band TL model \citep{JV2006} and phenomenology \citep{Allen13}, respectively. 

The good agreement over such a broad range of temperature \footnote{In Ref. \cite{SM} (section IV) we discuss how good fits for the k-integrated spectra are obtained here without the T-dependent $\alpha$ used in \cite{JV2006}.} justifies setting $\alpha_{PG}=\alpha_{TG}=0.6$, as defined above. These considerations explain the breaking of the scaling law, $\alpha_{PG}=\eta_{PG}$. 
The theory is presented for $v_{\rho -}=0.05 \AA eV$ (the only free parameter of this work). The gapped mode is slower $v_{\rho -}\ll v_\sigma, v_{\rho +}$ than the previously characterized single-band LL modes, which would explain weak anomalies in the specific heat at the temperature scale of the gap \cite{Chi_sChoi}.

\begin{figure}
    \includegraphics[width=40mm,height=40mm]{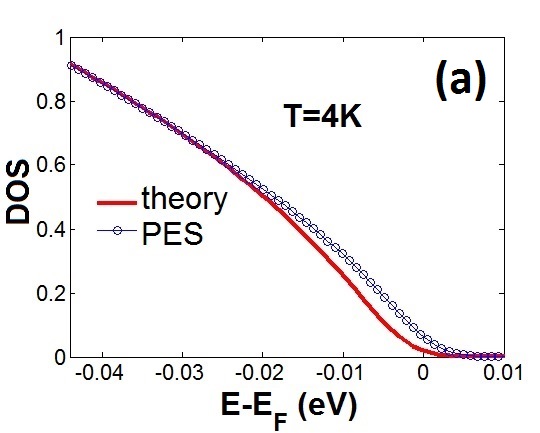}
    \includegraphics[width=40mm,height=40mm]{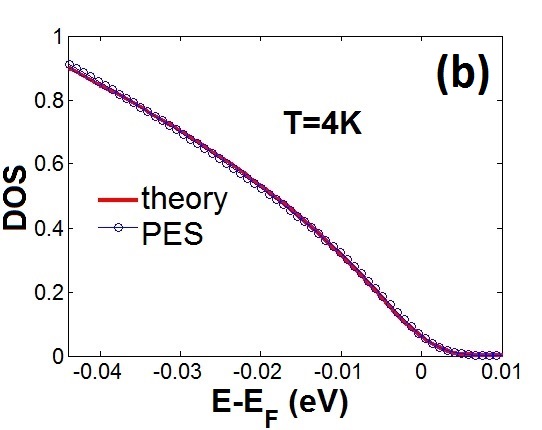}
    \includegraphics[width=40mm,height=40mm]{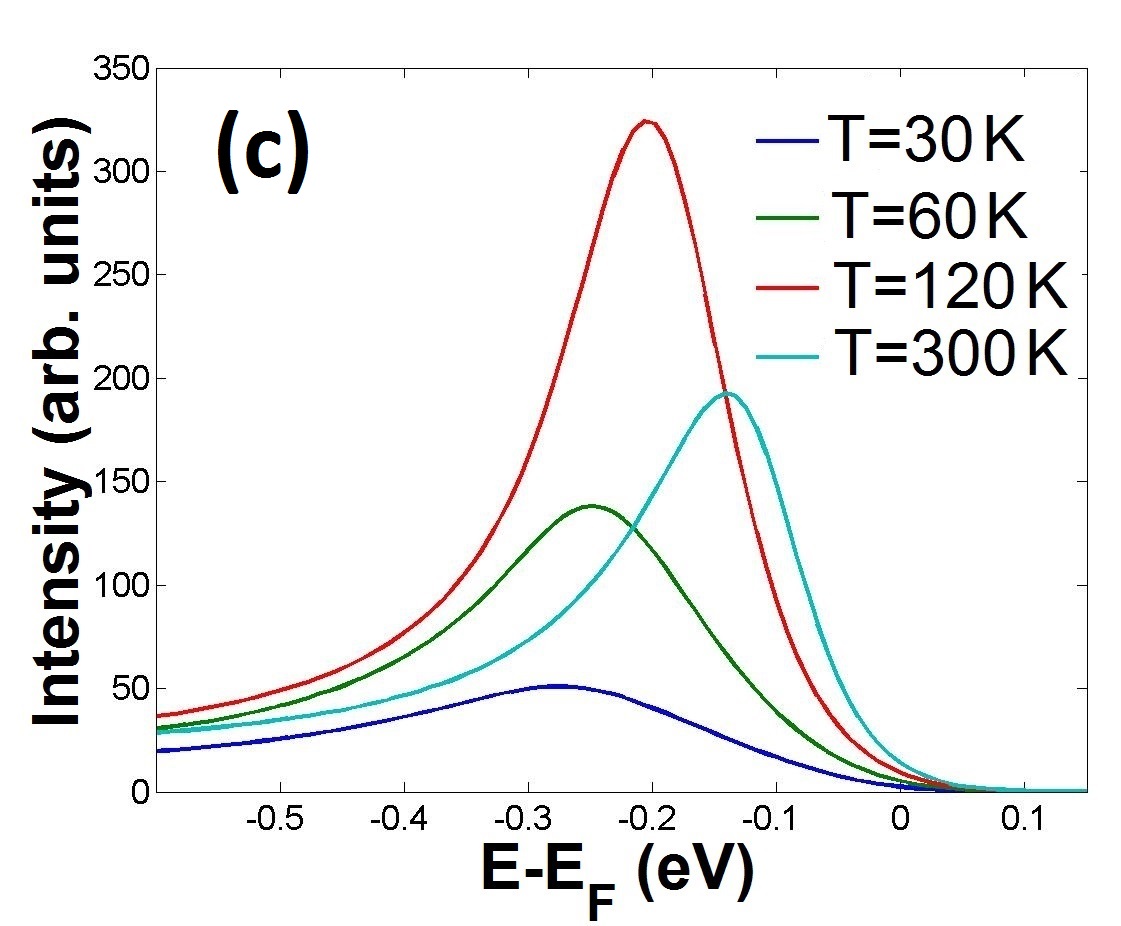}
    \includegraphics[width=40mm,height=40mm]{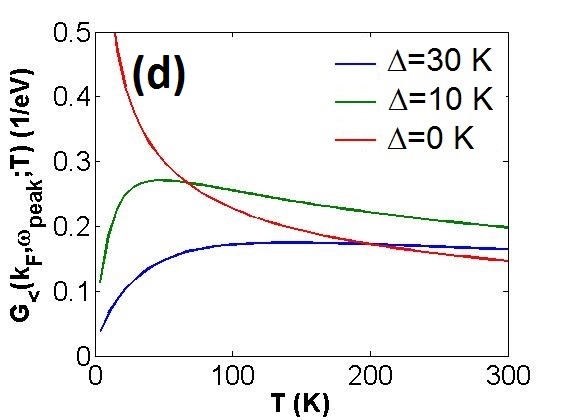}  
\caption {(Color online) (a,b) Comparison of previous single band TL fit \cite{Allen13} of 4K angle integrated data to gap mode theory (solid lines) (a) unshifted and (b) shifited by $3meV$. (c) Spectral functions at first order in perturbation theory at $k_F$ (d) Peak intensity against temperature. For a single LL ($\Delta=0$) we see $T^\eta$ behavior, on the contrary, a gapped spectral function does not diverge.}

\label{gap_implication}
\end{figure}

We now discuss the observability of the $\Delta=30K\approx 2.5meV$ $\rho -$ gap that we propose. For the two-particle optical conductivity, ungapped modes that couple to the photons,  especially the charged $\rho +$ mode, produce an ungapped response, consistent with optical spectroscopy \citep{Degiorgi} at $4K$ that has not found a CDW-type gap down to 1 meV.  In contrast, the k-integrated single-particle DOS should show a gap at $E_F$ because that gap energy must be overcome for electron removal or addition.  
 In our calculation, Fig. \ref{gap_implication} (a), the spectrum energy shift away from $E_F$ implied by the gap is clearly seen at low temperatures ($T<30$K).  However the best resolution of low T photoemission measurements to date ($5 meV$ \citep{Allen13}) is marginal for observing such a small gap and also a particularly precise calibration of $E_F$ in the measurement would be required.  As shown in  Fig. \ref{gap_implication} (b), an $E_F$ shift of $3 meV$ yields agreement between the new theory and experiment (as represented by its single-band TL fit \citep{Allen13}), but further experiment optimized to focus on this specific issue is required for a definitive test.

It is essential to discuss the role of single-particle hopping between the chains $t_\perp$ in this framework. 1D fluctuations can only suppress $t_\perp$ \cite{B_coupledLL}, but a gapped mode can render $t_\perp$ fully irrelevant. The instability towards a FL can be seen in the free propagator perpendicular to the chains, where the quasiparticle pole can develop \cite{B_coupledLL}:
\begin{equation}
G_{<Gap}^{(1)}=\frac{G_{<Gap}}{1-t_\perp(k_\perp) G_{<Gap}}
\label{1stOrderPerturbationTheory}
\end{equation}
This is a key point for the single-band LL theory which is unstable against FL physics at low T ($G_{<TL}(T=0)$ diverges). Direct evidence for FL crossover has never been found for LiPB \citep{Allen13}. In the gapped model at low-T, the holon peak shifts and decreases, Fig. \ref{gap_implication}(c). Therefore, $G_{<Gap}^{(1)}$ does not show any divergence for low enough $t_\perp$, and thus the robust q-1D behavior is intrinsic to the proposed model. In Fig. \ref{gap_implication}(d) we show the peak evolution in temperature for a pure LL and for two different values of the gap in this theory (keeping the value of all other parameters constant, see \cite{SM} for details). The qualitative features of the model are consistent with the observed LL signatures down to the superconducting transition.

As has been mentioned, we expect a transition from the non-Fermi liquid described here to a superconducting phase. There is evidence of possible triplet superconductivity \cite{mercure2012}, which is compatible with our model owing to the gapless spin mode \cite{Fradkin_Liu,GiamarchiBook,Chudzinski}.

\begin{figure}
        \includegraphics[width=80mm,height=30mm]{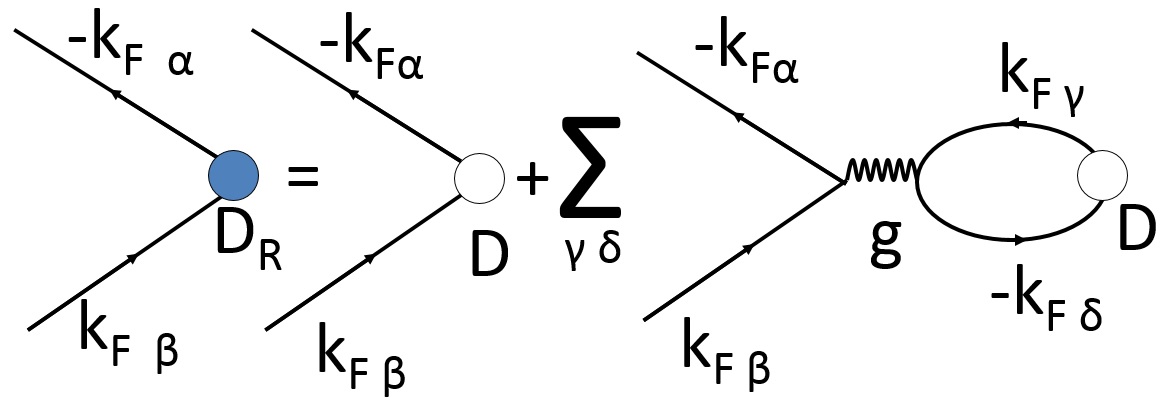}
\caption {(Color online)Renormalization of disorder due to interband coupling, g. Greek indices label spin degree of freedom.}
\label{diagram}
\end{figure}

The mysterious upturn in the resistivity is also compatible with the spectral theory proposed in this paper. Analogous to the single-band LL, finite resistivity arises from the interplay of two perturbations to the perfectly conducting fixed point: disorder and interband scattering. The latter plays in our model the same role as spin back-scattering plays for the single-band LL \cite{Giamarchi_Schulz}  i.e. promoting  the instability to the $\rho_{-}$ gapped phase increases  the sensitivity to disorder. This interplay is shown by a crossover in the resistivity \cite{Giamarchi_Schulz,Orignac_Giamarchi}) connecting one power law behavior at high-T to another at low-T, as actually observed in LiPB \cite{Allen13}.  So, rather than the activated behavior expected in a CDW transition, the LiPB upturn is actually associated with a mechanism of pinning driven  by instabilities in one-dimensional systems. The upward renormalization of the interband coupling affects the scattering (and resistivity) at second order, as shown in Fig. \ref{diagram}.  In other words, interband coupling increases the anomalous exponent making the disorder more relevant at the T-scale of the gap ($\Delta\sim30K$) and thus, the upturn is nearly independent of the amount of disorder in a given sample. 

	In summary, we propose that the many remarkable properties of LiPB can be understood in a unified way by considering the four boson modes implied by having two quasi-1d bands crossing $E_F$, specifically that the $\rho -$ mode acquires a small gap.  The spinless and chargeless gapped mode is not now directly observed,  but we show that its existence is signalled by the disagreements between ARPES data and single band LL theory, which we can now explain microscopically in detail, by the remarkably robust q-1D behavior of LiPB, which persists without FL crossover down in temperature to the onset of superconductivity, and plausibly by weak specific heat anomalies and a prominent powerlaw upturn of the resistivity at a temperature set by the gap.  The proposed triplet superconductivity is also compatible with the new theory. This general good success provides a strong motivation for new low temperature and higher resolution photoemission measurements specifically optimized to test for the small gap in the k-summed single-particle spectral function.

$\mathit{Acknowledgments}$: We thank Kai Sun and L. Dudy for fruitful discussions and insights.
This work has been funded by MINECO grants MINECO FIS2012-37549-C05-03 and FIS2015-64886-C5-5-P. NL acknowledges financial support from the
Spanish Ministry of Economy and Competitiveness, through
The “Mar\'ia de Maeztu” Programme for Units of Excellence in R\&D (MDM-2014-0377), and also hospitality from the University of Michigan where part of this work was done.

\bibliographystyle{ieeetr}
\bibliography{main_def1}

\end{document}